\documentclass[aip,pop,reprint]{revtex4-1}
\usepackage{graphicx}
\usepackage{ulem}
\usepackage{amssymb}
\usepackage{amsmath}
\usepackage{wasysym}
\usepackage{relsize}
\begin{document}


\newcommand{\be}{\begin{equation}}
\newcommand{\ee}{\end{equation}}
\newcommand{\bea}{\begin{eqnarray}}
\newcommand{\eea}{\end{eqnarray}}
\newcommand{\Tbar}{{\bar{T}}}
\newcommand{\En}{{\cal E}}
\newcommand{\K}{{\cal K}}
\newcommand{\U}{{\cal U}}
\newcommand{\GC}{{\cal \tt G}}
\newcommand{\Lop}{{\cal L}}
\newcommand{\DB}[1]{\marginpar{\footnotesize DB: #1}}
\newcommand{\q}{\vec{q}}
\newcommand{\kt}{\tilde{k}}
\newcommand{\Lopn}{\tilde{\Lop}}
\newcommand{\noi}{\noindent}
\newcommand{\ovn}{\bar{n}}
\newcommand{\ovx}{\bar{x}}
\newcommand{\ovE}{\bar{E}}
\newcommand{\ovV}{\bar{V}}
\newcommand{\ovU}{\bar{U}}
\newcommand{\ovJ}{\bar{J}}
\newcommand{\calE}{{\cal E}}
\newcommand{\calA}{{\cal A}}
\newcommand{\calC}{{\cal C}}
\newcommand{\ovphi}{\bar{\phi}}
\newcommand{\zt}{\tilde{z}}
\newcommand{\ttl}{\tilde{\theta}}
\newcommand{\nuv}{\rm v}
\newcommand{\ds}{\Delta s}
\newcommand{\fn}{{\small {\rm  FN}}}
\newcommand{\cc}{{\cal C}}
\newcommand{\cd}{{\cal D}}
\newcommand{\tth}{\tilde{\theta}}
\newcommand{\cb}{{\cal B}}
\newcommand{\cg}{{\cal G}}
\newcommand\norm[1]{\left\lVert#1\right\rVert}
\title{Enhanced space charge limited current for curved electron emitters}

\vskip 0.15 in

\author{Gaurav Singh}
\affiliation{
Bhabha Atomic Research Centre,
Mumbai 400 085, INDIA}
\affiliation{Homi Bhabha National Institute, Mumbai 400 094, INDIA}
\author{Raghwendra Kumar}
\affiliation{
Bhabha Atomic Research Centre,
Mumbai 400 085, INDIA}
\author{Debabrata Biswas}\email{dbiswas@barc.gov.in}
\affiliation{
Bhabha Atomic Research Centre,
Mumbai 400 085, INDIA}
\affiliation{Homi Bhabha National Institute, Mumbai 400 094, INDIA}


\begin{abstract}
  The maximum current that can be transported across a vacuum diode is limited by forces arising
  due to space charge. In a planar diode, the space charge limited (SCL) current
  density from an emitting patch is given by the Child-Langmuir (CL) law $J_{CL} \sim V_g^{3/2}/D^2$
  where $V_g$ is the potential difference across
  the diode and $D$ is the separation between the anode and cathode. 
  We show here analytically using the nonlinear line charge model
  that for a curved emitter in a planar diode configuration,
  the limiting current obeys the scaling
  relationship $J_{SCL} \sim \gamma_a V_g^{3/2}/D^2$
  where $\gamma_a$ is the apex field enhancement factor of the curved emitter.
  For an emitter
  with large height ($h$) to apex radius of curvature ($R_a$) ratio, the limiting current
  far exceeds the planar value. The result is
  verified using the particle-in-cell code PASUPAT for two curved emitters shapes.
\end{abstract}

\maketitle

\section{Introduction}

  The maximum current that can be transported across a vacuum diode is a quantity of immense interest
in vacuum micro-electronics, space navigation and especially in high-power vacuum electronic devices
such as  magnetron, gyrotron, vircator or relativistic backward wave oscillators \cite{benford}.
The limit on the current arises due to space charge
which give rise to an electric field that opposes the macroscopic
applied field in the diode. The current that enables an exact cancellation of these
forces is referred to as the space charge limited (SCL) current. It is a quantity that can
be reliably used to design systems and is used to model explosive emission \cite{benford} in
particle-in-cell (PIC) codes.

In a planar diode configuration, the SCL current density is given by the Child-Langmuir (CL) law \cite{child,langmuir}

\be
J_{CL} = \frac{4\epsilon_0}{9} \left( \frac{2e}{m} \right)^{1/2} \frac{V_g^{3/2}}{D^2}  \label{eq:CL}
\ee

\noi
where $e$ and $m$ are the electronic charge and mass respectively,
$V_g$ is the potential difference across  anode-cathode gap and $D$ refers
to the distance between them. Eq.~(\ref{eq:CL}) strictly holds for a diode where the
parallel plates are infinite in extent and the entire cathode participates in
emission. Such a situation may approximate the case when
the plate separation $D$ is very small compared to the size of the emitting region.
When they are comparable, a correction factor needs to be
incorporated \cite{luginsland96,lau,luginsland2002,zhang2017}

In recent decades, emission from curved surfaces under the application of an
electric field has been extensively researched. A curved emitter placed
perpendicular to the cathode plate in a planar diode configuration, has a local electric
field at the apex which can be expressed as $E_a = \gamma_a E_0$ where
$E_0 = V_g/D$ is the macroscopic field and $\gamma_a > 1$. This leads to
larger field emission (FE) currents at moderate fields and can hasten the transition
from cold-FE\cite{forbes2007} to thermal-FE\cite{jensen2006} and finally to  
explosive emission in high power systems\cite{benford,FE_relevance,forbes2008,jensen2015}.
In such situations of emission
from a curved surface, the space charge limited current is expected to
increase and its exact nature is a matter of considerable interest.
A clue that the scaling relationship $V_g^{3/2}/D^2$ should continue to hold for
curved emitters, follows from the general scaling law for
Schr\"odinger-Poisson system arrived at using dimensional analysis \cite{biswas_epl} wherein
$J \simeq \hbar^{3-2\alpha} V_g^\alpha/D^{5 - 2\alpha}$ where $\hbar = h/2\pi$, $h$ being
the Plank constant. For a purely classical
system $3-2\alpha = 0$ so that $J \simeq V_g^{3/2}/D^2$ irrespective of the
emitter shape\cite{other_combinations}.

The limiting current is known analytically for co-axial cylindrical diodes
and concentric spherical diode systems apart from the
planar (Cartesian) system \cite{zhang2017,LB23,LB24,zhu2013,darr2019}.
A similar treatment for a curved emitter in a planar diode configuration
seems beyond the scope of standard techniques and it is
therefore necessary to take recourse to approximate analytical methods and PIC
codes where space-charge limited emission from curved surfaces is modeled
adequately. The aim of this work is to establish an extension of the Child-Langmuir law for
curved emitters. This is especially important keeping in mind the fact that cathode
designs are often made using the vacuum electric field (in the absence of charges)
at the cathode, $E_C$. Thus the planar Child-Langmuir current density may be
expressed as\cite{zhang2017} $J_{CL} \sim E_C^{3/2}$ and its collisional counterpart,
the Mott-Gurney law as $J_{MG} \sim E_C^2$. Since the cathode field at the apex of a curved emitter
gets enhanced by a factor $\gamma_a$, this would imply a scaling $J_{CL} \sim \gamma_a^{3/2}$.
It is thus important to investigate whether such a scaling in fact holds.

\section{Approximate derivation of SCL current}

The space charge limited
current \cite{zhang2017,puri} for zero injection velocity corresponds to the condition
that the net electric field at the cathode is zero. We shall use approximate
methods which incorporates this basic requirement and highlights the physical aspects.
A simple derivation of the Child-Langmuir law based on vacuum capacitance and transit
time \cite{rose55,umstattd,zhang2017}
provides much physical insight. 

\subsection{The Planar Case}

Consider a planar diode with the anode and cathode
separated by a distance $D$ and having a potential difference $V_g$. The electric
field between the plates is $E_0 = V_g/D$ and the induced surface charge density on
the cathode plate is $\epsilon_0 E_0$. The total charge $Q_b$ induced on an area $\calA$
is thus $Q_b = \epsilon_0  \calA E_0 = \epsilon_0 \calA V/D$. This is also the magnitude of the charge
induced on the anode on an identical area $\calA$ since the field lines remain straight and 
perpendicular to the plates.

\begin{figure}[hbt]
  \begin{center}
    \vskip -0.75cm
\hspace*{0.35cm}\includegraphics[scale=0.375,angle=0]{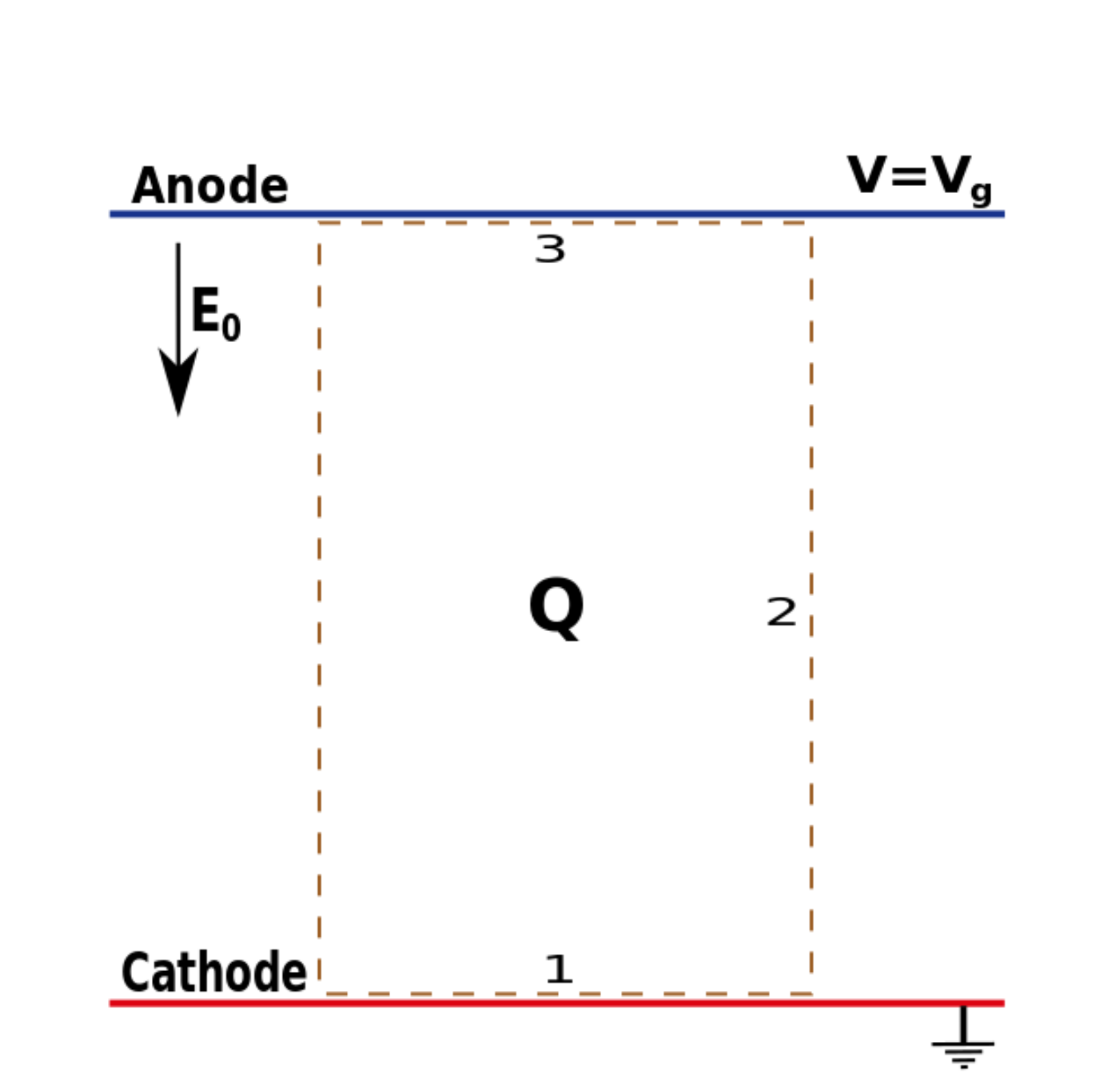}
\vskip -0.35cm

\caption{A schematic of a planar diode and a Gaussian surface marked by dashed lines}
\label{fig:gauss_planar}
\end{center}
\end{figure}

In order to address the question of space charge limited current, consider a charge filled
flux tube having a cross-sectional area $\calA$ and extending from the cathode ($z = 0$) to
the anode ($z = D$). 
Let the total charge contained be $Q$ such that the net field
(the macroscopic field $-E_0 \hat{z}$ and the space charge field) at the
cathode is zero. It is assumed here that the field lines remain straight and
perpendicular to the plates in the presence of the free charge $Q$ (see Fig.~\ref{fig:gauss_planar}).
Consider, a Gaussian surface covering this tubular volume (dashed lines), with the
faces parallel to the plates (marked 1 and 3) infinitesimally away from them. Since the net
field at the cathode is zero, $\int_1 E.dS = 0$ for the face (1) near the cathode. The
flux through the transverse surface (labeled 2) is also zero  while the
flux though the surface near the anode is ${E_0 \calA}$ assuming that the field
at the anode remains $E_0$ in the presence of the charge $Q$. In reality the field
at the anode $E_A = \alpha_1 E_0$ with $\alpha_1 > 1$. Thus, the
approximate total flux through the Gaussian surface is
$\oint E.dS =  E_A \calA \simeq E_0 \calA = \calA V_g/D = Q/\epsilon_0$
so that the free charge $Q = \epsilon_0 E_0 \calA = Q_b$ needs to reside in the
diode in order that the field at the cathode is zero.

\vskip -0.2in
Assuming now that the average transit time between the plates is $T_{av} = D/v_{av}$
where $v_{av} = v_{max}/2$ is the average speed and $v_{max} = (2e V_g/m)^{1/2}$
is the maximum speed, the space charge limited current through the diode is
$I_{SCL} = Q/T_{av} = \calA (\epsilon_0/2) (2e/m)^{1/2} V_g^{3/2}/D^2$.
Note that in the actual planar 1-D situation\cite{SCLpot,chauvin}, the average speed is
$v_{av} = v_{max}/\alpha_2$ where $\alpha_2 = 3$ while $\alpha_1 = 4/3$. Thus $\alpha_1/\alpha_2 = 4/9$
as in the Child-Langmuir law of Eq.~(\ref{eq:CL}) instead of $1/2$.
The approximate result is however close to the exact result and the derivation is
considerably simpler.

\subsection{SCL current for a curved emitter}
\label{subsec:SCL_curved}
\begin{figure}[hbt]
  \begin{center}
  \vskip -1.0cm
\hspace*{0.0cm}\includegraphics[scale=0.55,angle=0]{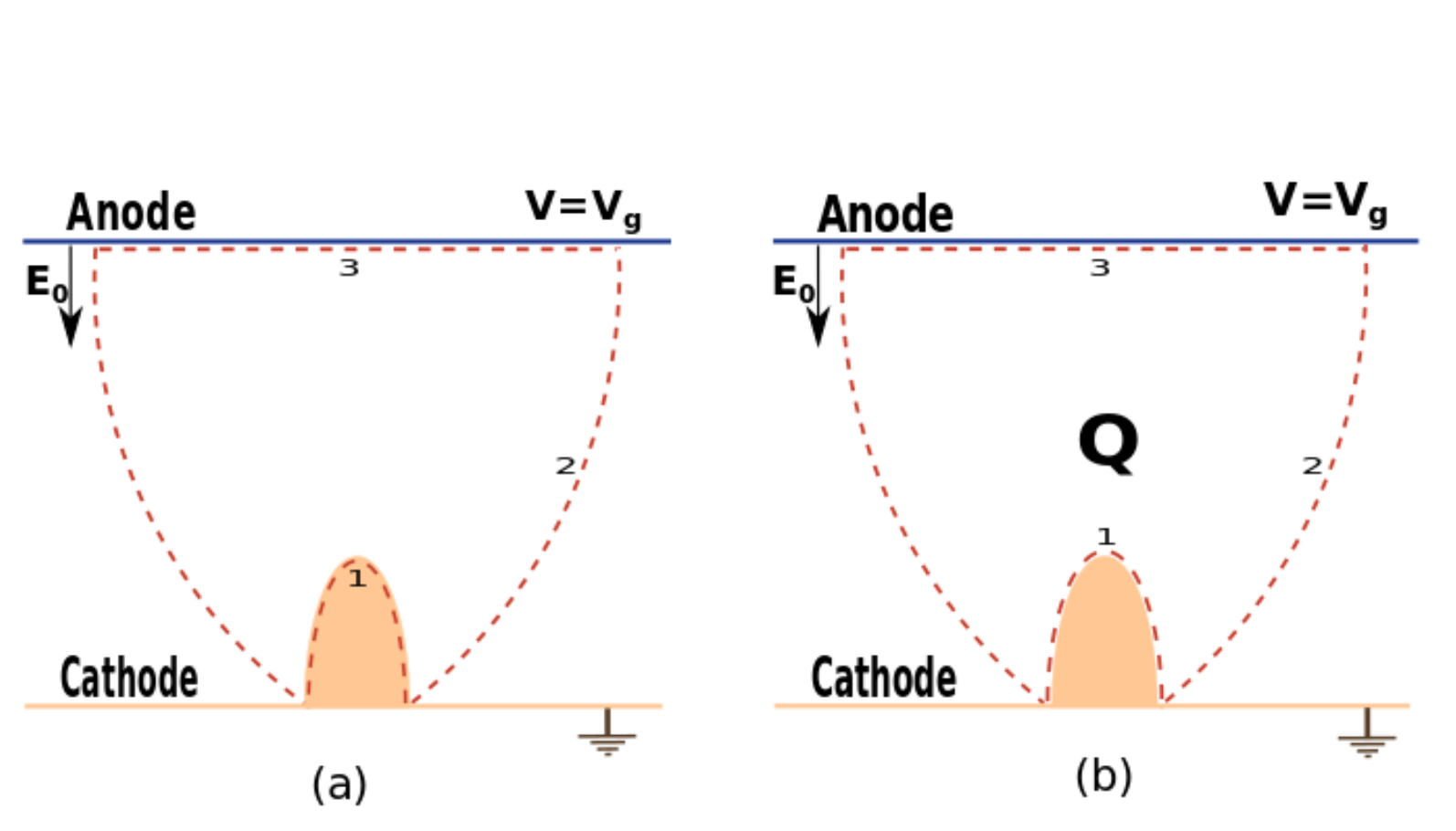}
\vskip -0.3cm
\caption{A schematic of a planar diode with a hemi-ellipsoidal emitter along with
  Gaussian surfaces marked by dashed lines.}
\label{fig:gauss_curved}
\end{center}
\end{figure}

\vskip -0.15in
We shall follow a similar procedure to determine the approximate SCL current for a
curved emitter. In order to keep the derivation simple, we shall consider a
hemi-ellipsoidal emitter of height $h$ and apex radius of curvature $R_a$
mounted on the cathode plane.
As before, consider the diode without any free charge. Let the charge induced
on the hemi-ellipsoidal emitter be $Q_b$. Consider a flux tube as a
Gaussian surface as shown in Fig.~\ref{fig:gauss_curved}a
consisting of surfaces 1,2 and 3. Since surface 1 is
inside the hemi-ellipsoid, the flux through this is zero. Further, surface 2
is aligned along the field lines and hence the flux is again zero. The flux
through surface 3 is $E_0 \calA_1$ where $\calA_1$ is the cross-sectional area
of the flux tube at the anode which is so far unknown. It follows from Gauss's law that
$E_0 \calA_1 = Q_b/\epsilon_0$.
Thus, $E_0 \calA_1$ can be determined if $Q_b$ can be calculated.

For a hemi-ellipsoid with the anode far away, the induced charge $Q_b$ can be
determined since the exact solution is known. It has also been established \cite{biswas_nonlinear}
that the surface charge density can be projected on the axis as a line charge
having linear density $\Lambda(z) = \lambda z$ where \cite{biswas_nonlinear,pogorelov}
$\lambda \simeq 4\pi\epsilon_0 E_0/[\ln(4h/R_a) - 2]$
for $h/R_a$ sufficiently large. The total bound charge therefore is
$Q_b = \int_0^h \lambda z~ dz = \lambda h^2/2$
Combining the expressions for $\lambda$ and $Q_b$, we have
$Q_b = \lambda h^2/2 = \gamma_a \pi\epsilon_0 E_0 R_a h $
where \cite{kosmahl,pogorelov,biswas_universal} $\gamma_a \simeq (2h/R_a)/[\ln(4h/R_a) - 2]$
is the field enhancement factor at the apex of the hemi-ellipsoidal emitter
(the local field at the apex is thus $E_0 \gamma_a$). Thus
$E_0 \calA_1 = \lambda h^2/2 = \gamma_a \pi E_0 b^2$
where we have used $h R_a = b^2$ where $b$ is the radius of the hemiellipsoid base.

We are now in a position to deal with the space charge limited current. Consider
the Gaussian surface shown in Fig.~\ref{fig:gauss_curved}b containing free charge $Q$.
In the SCL limit, the field on surface 1
is zero. We shall assume surfaces 2 and 3 to be identical to Fig.~\ref{fig:gauss_curved}a. Thus,
the flux through surface 2 is approximately zero while the flux through surface
3 is $E_0 \calA_1$. Using Gauss's law, the free charge 
$Q \simeq \epsilon_0 E_0 \calA_1  \simeq \gamma_a \epsilon_0 \pi E_0 b^2$.
Assuming as in the planar case that the average velocity $v_{av} = (2e V_g/m)^{1/2}/2$,
the SCL current for a curved emitter is

\vskip -0.1 in
\be
I_{SCL} \simeq (\pi b^2) \gamma_a \frac{\epsilon_0}{2} \left( \frac{2e}{m} \right)^{1/2} \frac{V_g^{3/2}}{D^2}.
\label{eq:SCL_curved}
\ee

\noi
Compared to a planar emitter of area $\calA = \pi b^2$, the SCL current
for a curved emitter is greater by a factor $\gamma_a$.

As in the planar case, there are a number of approximations that may alter the
factor $1/2$. These include the field at the anode,
the average transit time and finally the assumption that the $\calA_1$ does not
change in the presence of charges. The scaling $I_{SCL} \sim \gamma_a V_g^{3/2}/D^2$
should however be preserved despite these approximations.

An extension to a general axially symmetric curved emitter is somewhat non-trivial
but can be similarly carried out by noting that the projected line
charge density is in general nonlinear
and has the form \cite{biswas_universal} $\Lambda(z) = z f(z)$. Thus,

\vskip -0.15in
\be
Q_b  =  \int_0^h \Lambda(z) dz
 =  f(h) \frac{h^2}{2} \left[ 1 - \int_0^h \frac{z^2 f'(z)}{h^2 f(h)} dz \right]
\ee

\vskip -0.1in
\noi
where \cite{biswas_universal} $f(h) = 4\pi\epsilon_0 E_0 \gamma_a R_a/2h$.   
Thus, the total charge contained in the diode such that the field on cathode
vanishes is $Q \simeq \epsilon_0 E_0 \calA_1  = Q_b = f(h) (h^2/2) (1 - \calC)  = \gamma_a \pi E_0 R_a h~ (1 - \calC)$
where $\calC = \int_0^h (z^2 f'(z))/(h^2 f(h)) dz$
is a nonlinear correction factor that is a-priori unknown. Note that for a
hemi-ellipsoid, $f'(z) = 0$ so that $\calC = 0$.

Using the expression for average velocity as before, it follows that
$I_{SCL} \simeq (\pi R_a h) \gamma_a (\epsilon_0/2) (2e/m)^{1/2} (1 - \calC) V_g^{3/2}/D^2$
Thus, the linear dependence on the apex field enhancement factor $\gamma_a$ is expected for
other emitter shapes as well.

\subsection{Incorporating Anode-Proximity and Shielding}

The analysis so far has been for an isolated curved emitter.
We shall now consider two competing effects that have been ignored so far,
each of which affects the apex field enhancement factor.
The presence of the anode in close proximity to the emitter-apex alters the
line charge density. Thus the expression for $\lambda$ no longer holds and it is
necessary to include the effect of charges induced on the anode \cite{anodeprox}.
The net effect is an increase in local field at the emitter apex.

The presence of other emitters in close proximity also alters the local field
at the apex due to electrostatic shielding \cite{db_rudra1,rudra_db}. In such a situation,
the local field decreases compared to its isolated value. Finally, the presence
of other emitters coupled with the anode also contributes enormously\cite{db_rudra2,db_hybrid}.
Fortunately, all of these effects can be incorporated approximately
to express $\lambda$ as\cite{db_rudra2} 
$\lambda \simeq  4\pi\epsilon_0 E_0/[\ln(4h/R_a) - 2 - \alpha_A +  \alpha_{S} - \alpha_{{SA}}]$
where $\alpha_A$ accounts for
anode-proximity of an isolated emitter, $\alpha_S$ accounts for shielding due to
all other emitters and $\alpha_{{SA}}$ is the indirect effect of other emitters mediated
through the anode. The apex field enhancement factor thus takes the form
$\gamma_a \simeq (2h/R_a)/[\ln\big(4h/R_a\big) - 2 - \alpha_A +  \alpha_{S} - \alpha_{{SA}}]$
so that the relation
$\lambda \simeq 4\pi\epsilon_0 E_0 \gamma_a R_a/2h$
continues to hold. Similarly the expression for $f(h)$ applies for a nonlinear line-charge
in the presence of other emitters and the anode. Thus, the central results for
SCL current arrived at earlier in this section, continue to hold.

\section{Results for curved emitters using PIC}
\label{sec:results}

\subsection{Comparison with an existing result}

In [\onlinecite{zhu2015}], the
localized current density at the apex of a
hyperboloid was reported to scale as $V_g^{3/2}/{D^m}$ with $m = 1.1 - 1.2$.
This appears to be at variance with the results derived here and therefore needs scrutiny.

A diode in [\onlinecite{zhu2015}] consists of two hyperboloids, one of which is a plane
at $z = 0$ acting as an anode and the other (cathode) with its apex located at $z = D$
having an apex radius of curvature $R_a$. The local field at the apex is
$E_a = 2V_g/R_a/\ln(4D/R_a) = [(V_g/D) 2D/R_a]/\ln(4D/R_a)$
for $D >> R_a$ so that the enhancement factor can be expressed as 
$\gamma_a \simeq (2D/R_a)/\ln(4D/R_a)$.

The regime explored in [\onlinecite{zhu2015}] is $D \in [500-1000]$nm with $R_a = 50$nm
and $100$nm. For $R_a = 50$nm, the apex enhancement factor can be approximated
by the fit $\gamma_a \simeq 0.952 (D/R_a)^{0.754}$. Thus, the scaling
$I_{SCL} \sim \gamma_a V_g^{3/2}/D^2 \sim V_g^{3/2}/D^{2 - 0.754} = V_g^{3/2}/D^{1.246}$.
The exponent of $D$ is close to the value of $m$ observed in [\onlinecite{zhu2015}]. 

\subsection{Comparison using the PIC code PASUPAT}

In order to further validate the analytical findings reported here, we shall
use a three dimensional fully electromagnetic relativistic PIC code\cite{Birdsall} named PASUPAT
developed by the authors. It uses the Yee grid based
Finite Difference Time Domain Method to solve Maxwell's equations in electromagnetic
solver module \cite{yee,Taflove,Inan_Marshal}.
The code also has an electrostatic solver, which currently uses the multigrid method\cite{multigrid}
to solve Poisson equation. The charge-conserving current weighting scheme of Esirkepov\cite{esirkepov}
is used to assign current densities to the grid points and the standard Boris method\cite{Birdsall}
is employed to move the charged particles. Apart from the Dirichlet and Neumann boundary conditions,
a variety  of open boundary conditions have been incorporated
both in the electromagnetic and electrostatic \cite{POP_es_abc} modules. 
For handling curved surfaces, the electromagnetic solver uses the the Dey-Mitra algorithm\cite{DM,Benkler}
while for the electrostatic solver, the discretization scheme in the cut-cell (computational cell lying
partially inside two medium) has been modified to account for the reduced spacing between mesh
points\cite{cut-cell-ES}. For space-charge limited emission from curved surfaces,
we have adapted the  algorithms
presented in [\onlinecite{SCL_LL}] and [\onlinecite{loverich}]. The code uses VTK \cite{vtk} library for
writing files for visualization of simulation data.
PASUPAT has been tested on standard benchmark problems. In the present context, it reproduces the
Child-Langmuir law for planar diode, both for the finite and infinite emission area. We shall use it
here to study the space charge limited current for curved emitters in a planar
diode configuration.

We consider space charge limited emission from (a) a hemi-ellipsoid emitter
and (b) a hemi-ellipsoid on a cylindrical post (HECP). The anode, cathode and
emitter are assumed to be perfect electric conductors. Periodic boundary condition imposed on the
side walls of the simulation domain. Typical anode-cathode gap $D$ was taken to be $15\mu$m
while the number of computational cells along direction of propagation of beam,
$N_z$, was typically 256 or 512,
while in the transverse direction $N_x$ and $N_y$ were taken to be 128.
We have checked for convergence against $N_x$, $N_y$, $N_z$ and time step $\Delta t$. The total number
of macro-particles is typically $1.5\times 10^5$.

\begin{figure}[hbt]
  \begin{center}
    \vskip -0.5cm
\hspace*{-.5cm}\includegraphics[scale=0.325,angle=0]{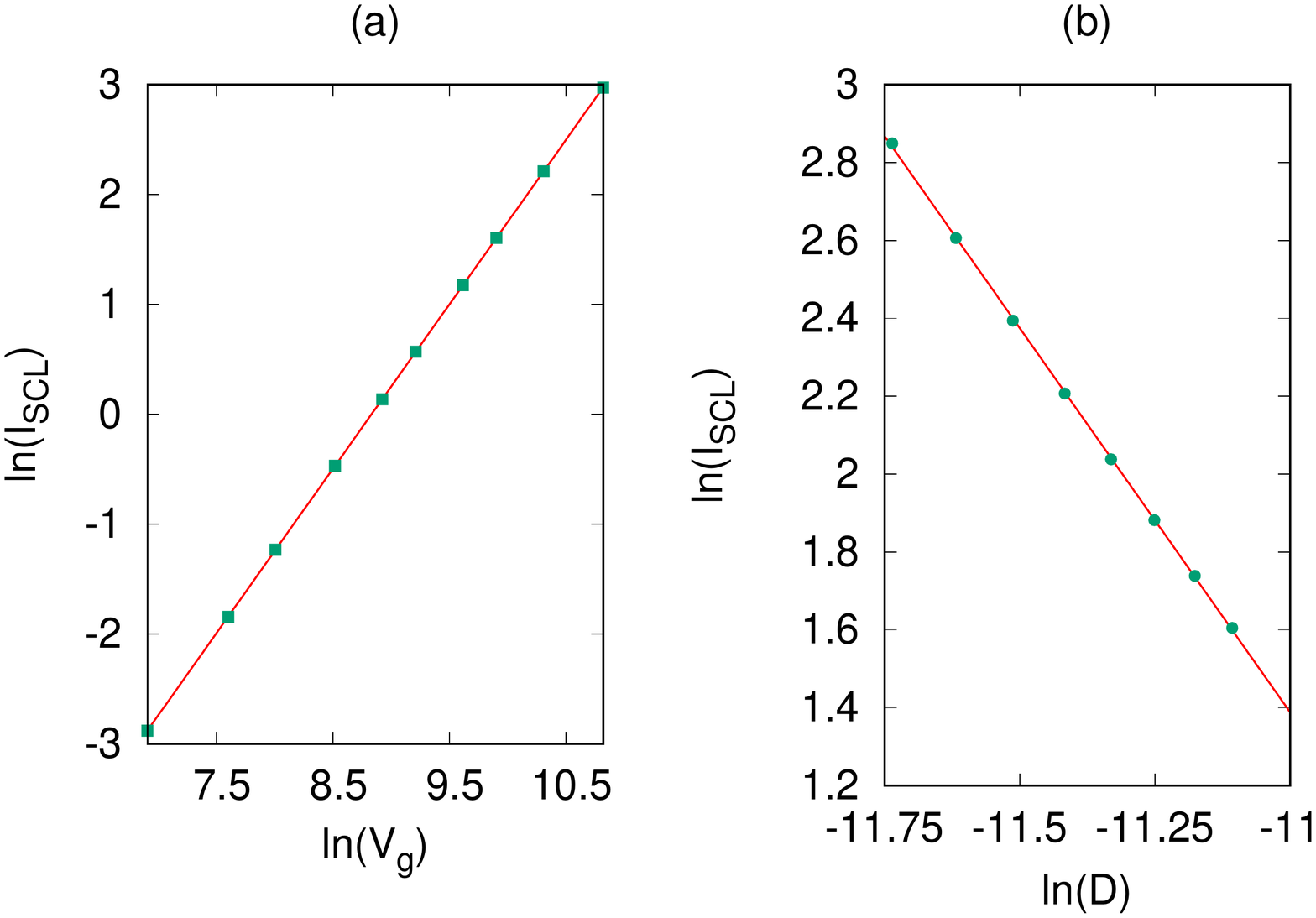}
\vskip -0.6cm
\caption{Verification of scaling with $V_g$ and $D$. Points are results obtained from simulation
  and line is best fit to the PIC simulation data. The emitter height $h = 3\mu$m.
  (a) The anode-cathode gap is held fixed at $D = 15\mu$
  while $V_g$ is varied. Slope of the line is $\approx 1.49$ (b)
  The voltage is held fixed at $V_g = 30$kV while $D$ is varied. The slope of the fitted line is $\approx 1.98$.
  }
\label{fig:IV_plot_fit}
\end{center}
\end{figure}

Fig.~\ref{fig:IV_plot_fit} shows the voltage scaling for an HECP emitter
in a parallel plate diode configuration. The base
radius $b = 1.5\mu$m, the height of the post is $1\mu$m while the height of the hemiellipsoid-cap is
$3\mu$m. We plot in Fig.~\ref{fig:IV_plot_fit}a, $\ln(I_{SCL})$ against $\ln(V_g)$. Fitting a straight line yields a slope of 1.49
which is close to analytical prediction $I_{SCL} \sim V_g^{3/2}$.
The scaling with anode-cathode plate gap $D$ is shown in
Fig.~\ref{fig:IV_plot_fit}b (right panel) with $V_g = 3$kV.
Fitting a straight line to the $\ln(I_{SCL})$ against $\ln(D)$ yields a slope of -1.98. These results further
validate the voltage ($V_g^{3/2}$) and anode-cathode gap ($D^{-2}$) scaling for curved emitters arrived at
analytically. A similar scaling has
been found to hold for the hemi-ellipsoid emitter.

\begin{figure}[hbt]
  \begin{center}
    \vskip -0.35in
\hspace*{-.5cm}\includegraphics[scale=0.325,angle=0]{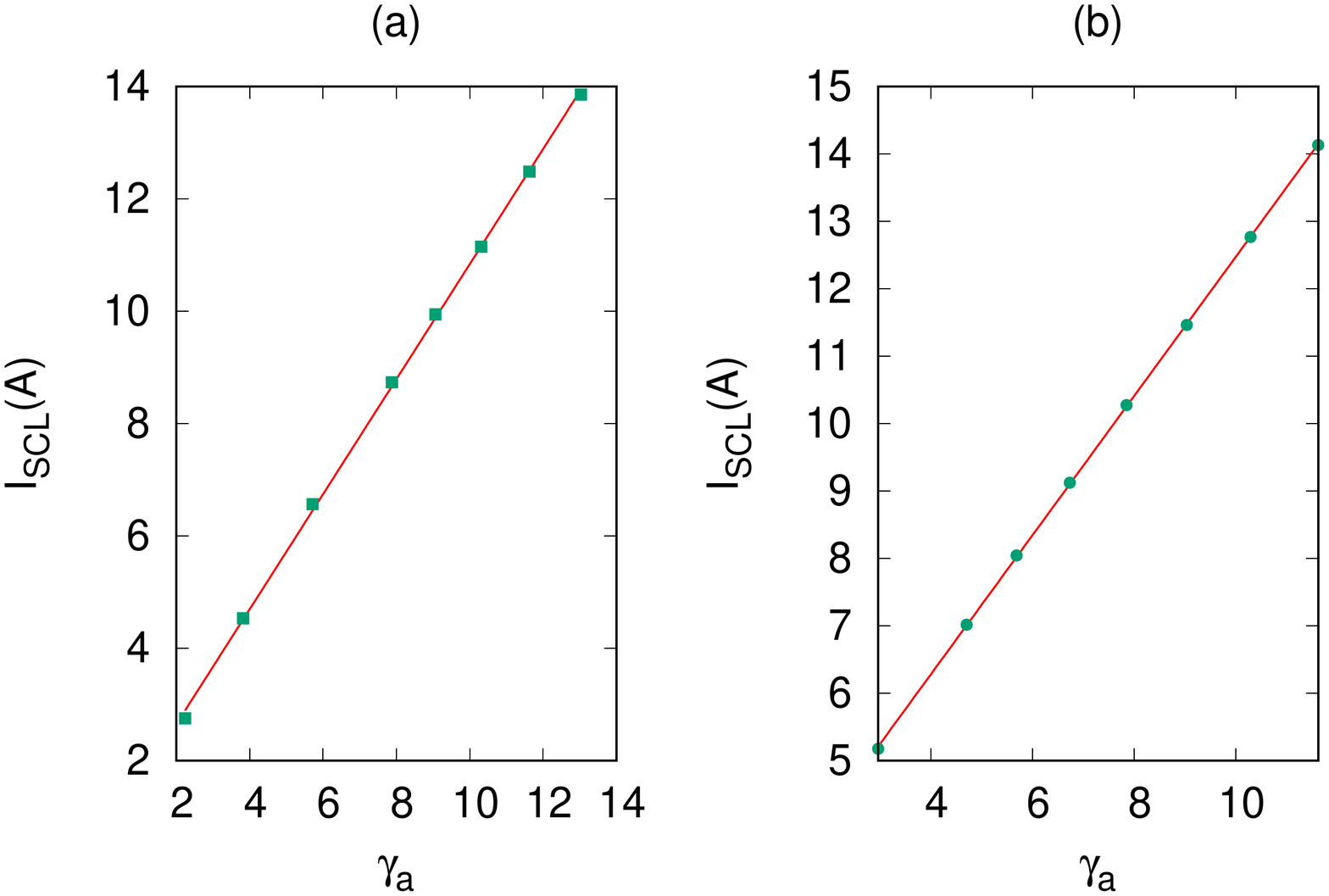}
\vskip -.75cm
\caption{Scaling of $I_{SCL}$ with the apex enhancement factor $\gamma_a$ for (a) hemiellipsoid
  (b) a hemiellipsoid endcap on cylindrical post. In both cases, the
  anode-cathode gap is $D = 15\mu$m and the base radius of the emitter $b = 1.5\mu$m.
  The straight line is the best
  fit while the points are PIC simulation data.
  In (b) the height of the endcap is increased
  keeping the cylinder fixed.}
\label{fig:gamma_vs_I}
\end{center}
\end{figure}

Finally, we present our study of variation of anode current with the apex field enhancement factor $\gamma_a$ in
Fig.~\ref{fig:gamma_vs_I}. The left panel (Fig.~\ref{fig:gamma_vs_I}a)
shows the result for a hemiellipsoid.  The enhancement factor is changed by
varying the height of the hemiellipsoid while keeping the base radius fixed.
The solid points are results obtained from PIC simulation while the straight
line is best fit. It is clear that as predicted by our theory, anode current scales
linearly with $\gamma_a$.

\vskip -0.25in
We next consider 
an emitter having a hemi-ellipsoid end-cap mounted on top of a cylindrical post, a clear case of a
nonlinear line charge density. As in the simulation study presented above, we change the height
of emitter to change the apex field enhancement factor.   
The space charge limited current for different $\gamma_a$ is plotted in the right panel
(Fig.~\ref{fig:gamma_vs_I}b). As before, solid points represent PASUPAT simulation data and
line is best fit. It is evident again that the SCL current scales linearly with $\gamma_a$.
We have tested these results for other geometries as well.

In conclusion, we have established that the space charge limited current for curved emitters obeys
the scaling relationship $I_{SCL} \sim \gamma_a V_g^{3/2}/D^2$ which reduces to the Child-Langmuir law for $\gamma_a = 1$.
Importantly, the scaling with $\gamma_a$ is linear and not $\gamma_a^{3/2}$ or $\gamma_a^2$ that
a straightforward extension of the planar law would imply.\\

{\it Data Availability}: The data that supports the findings of this study are available within the article.\\

{\it Acknowledgements}~\textemdash~The authors acknowledge the contributions of Vibhuti Duggal and Kislay Bhatt in
the parallelization of PASUPAT and thank Computer Division, BARC for discussions and valuable support.

\vskip -0.75 in
\section{References} 
\vskip -0.2in


\begin{thebibliography}{99}
\bibitem{benford} J. Benford, J. A. Swegle, and E. Schamiloglu, {\it High Power Microwaves},
 Taylor \& Francis Group,  New York (2007).
\bibitem{child} C.~D.~Child, Phys. Rev. 32, 492 (1911).
\bibitem{langmuir} I.~Langmuir, Phys. Rev. 2, 450 (1913).
\bibitem{luginsland96} J.~W.~Luginsland, Y.~Y.~Lau, and R.~M.~Gilgenbach, Phys. Rev. Lett. 77, 4668 (1996).
\bibitem{lau} Y.~Y.~Lau, Phys. Rev. Lett. 87, 278301 (2001).
\bibitem{luginsland2002} J.~W.~Luginsland, Y.~Y.~Lau, R.~J.~Umstattd, and J.~J.~Watrous, Phys. Plasmas 9, 2371 (2002).
\bibitem{zhang2017} P.~Zhang, A.~Valfells, L.~K.~Ang, J.~W.~Luginsland, and Y.~Y.~Lau, Appl. Phys. Rev. 4, 011304 (2017).
\bibitem{forbes2007} R.~G.~Forbes and J.~H.~B.~Deane, Proc. R. Soc. A.463, 2907 (2007).
\bibitem{jensen2006} K.L.Jensen, Appl. Phys. Lett. 88, 154105 (2006).
\bibitem{FE_relevance} The CL current density is not directly relevant for pure
  field emission since the electrostatic field at the cathode is not driven to zero by
  the space charge, at least in planar cases\cite{forbes2008, jensen2015}.
  The SCL analysis, originally made for thermionic emission, is relevant in situations
  where an `infinite' source of electrons is present, such as due to the formation of
  a cathode plasma in explosive emission\cite{benford}.
\bibitem{forbes2008} R.~Forbes, J.~Appl.~Phys., 104, 084303 (2008).
\bibitem{jensen2015} K.~L.~Jensen, D.~A.~Shiffler, I.~M.~Rittersdorf, J.~L.~Lebowitz, J.~R.~Harris, Y.~Y.~Lau, J.~J.~Petillo, W.~Tang, and J.~W.~Luginsland, J. Appl. Phys. 117, 194902 (2015).
\bibitem{biswas_epl} D.~Biswas and R.~Kumar, EPL 102, 58002 (2013).
\bibitem{other_combinations} For a diode with other length scales apart from $D$, the denominator
can have quadratic combinations other than $D^2$. 
\bibitem{LB23} I.~Langmuir and K.~B.~Blodgett, Phys. Rev. 22, 347 (1923).
\bibitem{LB24} I.~Langmuir and K.~B.~Blodgett, Phys. Rev. 24, 49 (1924).
\bibitem{zhu2013} Y.~B.~Zhu, P.~Zhang, A.~Valfells, L.~K.~Ang, and Y.~Y.~Lau, Phys. Rev. Lett. 110, 265007 (2013).
\bibitem{darr2019} A.~M.~Darr and A.~L.~Garner, Appl. Phys. Lett. 115, 054101 (2019). 
\bibitem{puri} R.~R.~Puri, D.~Biswas, and R.~Kumar, Phys. Plasmas 11, 1178 (2004). 
\bibitem{rose55} A.~Rose, Phys. Rev. 97, 1538 (1955).
\bibitem{umstattd} R.~J.~Umstattd, C.~G.~Carr, C.~L.~Frenzen, J.~W.~Luginsland and Y.~Y.~Lau,
  American Journal of Physics 73, 160 (2005).
\bibitem{SCLpot} The space-charge limited electrostatic potential for a parallel plate diode is $V(z) = V_g~ (z/D)^{4/3}$.
  It satisfies the Poisson equation $d^2 V/dz^2 = \rho(z)/\epsilon_0$ where $\rho(z) = J_{CL}/\sqrt{2eV(z)/m}$. 
  See for instance Eq.~(55) of [\onlinecite{chauvin}]. It follows that $\alpha_1 = 4/3$ and $\alpha_2 = 3$.
\bibitem{chauvin} N.~Chauvin, `Space-charge effect', DOI: 10.5170/CERN-2013-007.63, https://arxiv.org/abs/1410.7991
\bibitem{biswas_nonlinear} D.~Biswas, G.~Singh, and R.~Kumar, J.~Appl.~Phys. 120, 124307 (2016).
\bibitem{pogorelov} E.~G.~Pogorelov, A.~I.~Zhbanov, and Y.~C.~Chang, Ultramicroscopy 109, 373 (2009).
\bibitem{kosmahl} H.~G.~Kosmahl, IEEE Trans. Electron Devices 38, 1534 (1991).
\bibitem{biswas_universal} D~ Biswas, Phys. Plasmas 25, 043113 (2018).
\bibitem{anodeprox}  D.~Biswas, Physics of Plasmas, 26, 073106 (2019).
\bibitem{db_rudra1} D.~Biswas and R.~Rudra, Physics of Plasmas 25, 083105 (2018).
\bibitem{rudra_db} R.~Rudra and D.~Biswas, AIP Advances, 9, 125207 (2019).
\bibitem{db_rudra2}  D.~Biswas and R.~Rudra, J.~Vac.~Sci.~Technol.~B, J. Vac. Sci. Technol. B, 38, 023207
  (2020).
\bibitem{db_hybrid} D.~Biswas, preprint, https://arxiv.org/abs/2005.05700
\bibitem{zhu2015} Y.~B.~Zhu and L.~K.~Ang, Physics of Plasmas, 22, 052106 (2015).
\bibitem{Birdsall} C K Birdsall and A. B. Langdon, ``Plasma Physics via Computer Simulation", New York: McGraw-Hill, 1985.
\bibitem{yee} K.S. Yee, IEEE Trans. Antennas Propagat., vol. AP-14, No. 3, pp. 302-
307, 1966.
\bibitem {Taflove} A. Taflove and S. C. Hagness, ``Computational Electrodynamics", second Edition,  Artech House, Boston, 2000.
\bibitem {Inan_Marshal} U. S. Innan and R. A. Marshal, ``Numerical Electromagnetics: The FDTD Method", Cambridge University Press, Cambridge UK, 2011.
\bibitem{multigrid} W.~L.~Briggs, V.~E.~Henson, and S.~F.~McCormick, ``A Multigrid Tutorial, 2nd Edition",  SIAM, ISBN: 978-0-89871-462-3, (2000).
\bibitem{esirkepov} T.~Zh.~Esirkepov, Comput. Phys. Comm. 135, 144-153 (2001).
\bibitem{POP_es_abc} D. Biswas, G. Singh, and R. Kumar Physics of Plasmas 22, 093119 (2015).
\bibitem{DM} S.~Dey and R.~Mittra, IEEE Microwave and Guided Wave Letters, VOL. 7, NO. 9, 1997.
\bibitem{Benkler}S. Benkler, N. Chavannes and N. Kuster, IEEE Transactions on Antennas and Propagation, 54, 1843 (2006).
\bibitem{cut-cell-ES} L.~N.~Dworsky, ``Introduction to Numerical Electrostatics Using MATLAB"  John Wiley \& Sons, Inc. (2014)
\bibitem{SCL_LL}J. J. Watrous, J. W. Luginsland, and G. E. Sasser, Physics of Plasmas 8, 289 (2001).
\bibitem{loverich} J.  Loverich,  C.  Nieter,  D.  Smithe,  S.  Mahalingam,  and  P.  Stoltz, ``Charge conserving emission from conformal boundaries in electromagnetic PIC simulations", (2009) Available at 
https://www.researchgate.net/profile/John\_Loverich/publication/
\bibitem{vtk} W. Schroeder, K.~Martin and B.~Lorensen, ``The Visualization Toolkit'', 4rd Edition. Kitware, (2006), ISBN 978-1-930934-19-1. Available at url:  http://www.vtk.org

\end{thebibliography}
\end{document}